\documentclass[onecolumn]{aa}  

\usepackage{graphicx}
\usepackage{txfonts}
\begin{document}

   \title{A filament eruption on the RS~CVn-type star UX~Ari}
   %\subtitle{I. Overviewing the $\kappa$-mechanism}

   \author{Dongtao Cao
          \inst{1,2,3}
          \and
          Shenghong Gu\inst{1,2,4}
          }

   \institute{Yunnan Observatories, Chinese Academy of Sciences, Kunming 650216, China\\
              \email{dtcao@ynao.ac.cn, shenghonggu@ynao.ac.cn}
         \and
             Key Laboratory for the Structure and Evolution of Celestial Objects, Chinese Academy of Sciences, Kunming 650216, China
         \and
             International Centre of Supernovae, Yunnan Key Laboratory, Kunming 650216, China
         \and
             School of Astronomy and Space Science, University of Chinese Academy of Sciences, Beijing 101408, China
             }
   \authorrunning{Cao \& Gu}
%  \date{Received September 15, 1996; accepted March 16, 1997}

% \abstract{}{}{}{}{}
% 5 {} token are mandatory
  \abstract
   {Stellar flares can be accompanied by filament/prominence eruptions, which may lead to stellar coronal mass ejections (CMEs) that have significant impacts on surrounding exoplanets and stellar evolution.}
   {We report the capture of a filament eruption associated with an optical flare on the RS CVn-type star UX Ari.}
   {Based on high-resolution spectroscopic observations, we investigate the Doppler-shifted absorption features appeared in the H$_{\alpha}$ flaring spectra, which can be indicative of a filament eruption.}
   {A huge optical flare was observed, which lasted for a duration of at least 150~hours and released an energy of about $1.47~\times~10^{37}$~erg in the H$_{\alpha}$ line. The corresponding bolometric white-light energy is converted to be $1.03 \times 10^{39}$~erg, indicating that the flare is comparable to the most intense stellar superflares. A blueshifted absorption feature with a bulk velocity of -143.9~km~s$^{-1}$ emerged in the H$_{\alpha}$ line region during the initial phase of the flare, which then decelerated and turned into a redshifted absorption. This suggests that a filament eruption happened and ultimately fell back toward the star, suggesting that it represents a failed filament eruption. The erupted filament mass of $8.5-15.0~\times~10^{20}$~g and the kinetic energy of $8.8-15.5~\times~10^{34}$~erg are derived. Furthermore, the relationship between the filament eruption and the flare is likely characterized by a sympathetic eruption.}
   {}


   \titlerunning{A filament eruption on UX~Ari}
   \maketitle
   \nolinenumbers
%_________________________________________________________
\section{Introduction}
Solar flares can be accompanied by filament (also known as a prominence when viewed on the solar limb) eruptions and coronal mass ejections (CMEs), all of which have significant impacts on space weather \citep{Forbes2000, Koleva2021}. Stellar superflares (with energies of 10$^{33}$--10$^{38}$~erg; \citealt{Schaefer2000}) have been frequently observed on cool stars, particularly in the surveys conducted by the Kepler space telescope and the Transiting Exoplanet Survey Satellite (TESS) \citep[e.g.][]{Maehara2012, Tu2020, Yang2023}. Superflares may be associated with the production of much larger CMEs, which may affect the potential habitability of the  surrounding exoplanets \citep{Airapetian2016, Cherenkov2017, Hazra2022} and contribute significantly to mass and angular momentum losses during stellar evolution \citep{Aarnio2012, Osten2015}. 

In the solar case, eruptive filaments/prominences are commonly believed to form the CME cores in frequently observed three-part structure of solar CMEs \citep{Forbes2000}. Consequently, stellar filament/prominence eruptions can be indirect evidences of stellar CMEs and its investigation may also be an effective means to obtain information on formation and evolution of stellar CMEs in lower coronae. 

The detection of Doppler-shifted absorption and emission signatures in chromospheric line profiles provides one way to search for stellar filament/prominence eruptions and potential CMEs. Currently, there have been some such observational cases on cool stars \citep[e.g.,][]{Houdebine1990, Vida2019, Koller2021, Namekata2022, Wu2022, Lu2022, Inoue2023, Cao2024, Namekata2024}. Among these cases, only one eruptive filament had been detected on the young solar-type star EK~Dra \citep{Namekata2022} and a CME is probably associated with the eruption.

UX~Ari (=~HD~21242~=~BD+28$^{\circ}$532) is a non-eclipsing spectroscopic binary system, consisting of a K0~IV primary star and a G5~V secondary one, with an orbital period of about 6.44 days. UX~Ari is a highly active RS~CVn-type star. Its K0~IV primary component shows significant starspot activity and strong chromospheric emission in the H$_{\alpha}$, $\mbox{Ca~{\sc ii}}$ H~\&~K, and $\mbox{Ca~{\sc ii}}$ IRT lines. In addition, UX~Ari is also a high-rate flaring star, exhibiting flares several times in a wide range of wavelength \citep{Simon1980, Elias1995, Montes1996, Franciosini2001, Gu2002, Richards2003, Cao2017, Kurihara2024}. This indicates that UX~Ari may have frequent filament/prominence eruptions and CMEs based on the close association between solar CMEs and energetic flares.

This study focuses on a filament eruption that happened during the initial phase of a huge optical flare on UX~Ari. In Sect.~\ref{sec2}, we provide details of the observations, data reduction, and spectral subtraction of chromospheric lines. A huge optical flare and analysis of the flaring spectra are described in Sect.~\ref{sec3}, and the filament eruption is discussed in Sect.~\ref{sec4}. In Sect.~\ref{sec5}, we conclude and summarize the new results.
%%%%%%%%%%%%%%%%%%%%%%%%%%%%%%%%%%%%%%%%%%%%%%%%%%%%%%%%
%-------------------------------------------------------------------------------------------------------------------------%
\begin{figure}
\centering
\includegraphics[width=7.5cm,height=6.0cm]{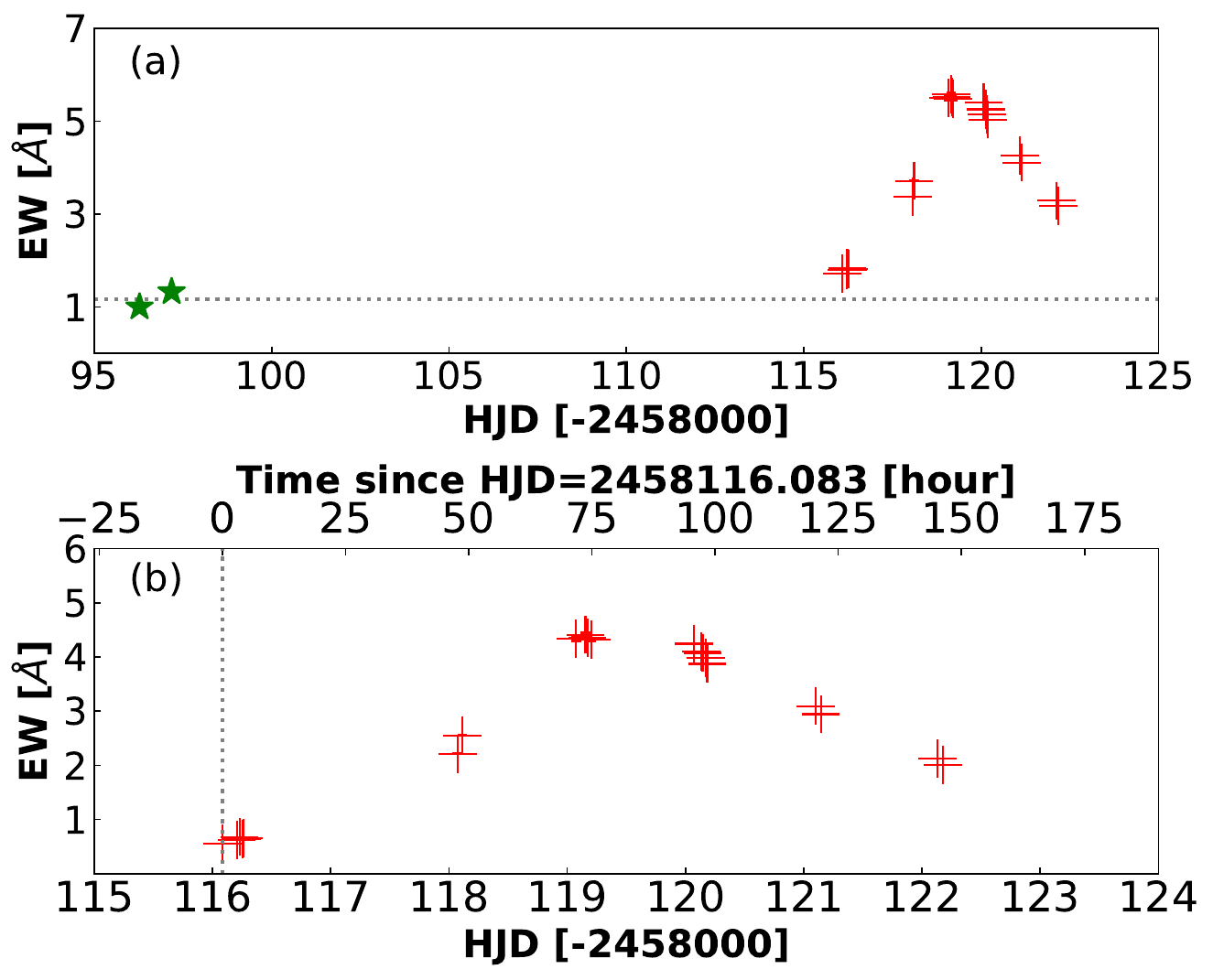}
\caption{Temporal variation of the EWs of the H$_{\alpha}$ line. (a) The EWs of the subtracted H$_{\alpha}$ spectra plotted against HJD. The dotted line represents the intrinsic chromospheric emission level. (b) The EWs subtracted the intrinsic chromospheric emission, plotted against HJD and flare time. The dotted line indicates zero point of the flare time.}
\label{Fig1}
\end{figure}
%-------------------------------------------------------------------------------------------------------------------------%
\section{Observations, data reduction, and spectral subtraction}\label{sec2}
Spectroscopic observations of UX~Ari were obtained by using the fiber-fed high-resolution spectrograph (HRS) installed on the 2.16m telescope at the Xinglong station of the National Astronomical Observatories, Chinese Academy of Sciences, China in 2017/2018 \citep{Fan2016}. HRS produces spectra with a resolving power of R~=~$\lambda$/$\Delta\lambda$~$\simeq$~48000 in a wavelength range of 3900--10000~\AA, using a $4096 \times 4096$ pixel CCD detector. The detailed observing information is listed in Table~\ref{taba1}.

The data are reduced with the IRAF\footnote{IRAF is distributed by the National Optical Astronomy Observatories, which is operated by the Association of Universities for Research in Astronomy (AURA), Inc., under cooperative agreement with the National Science Foundation.} package, following the prescription in \citet{Cao2024}. Moreover, the heavy telluric lines appeared in the chromospheric lines are removed by using a telluric template derived from the spectrum of the rapidly rotating early-type star HR~8858 (B5~V, $vsini$ = 316~km~s$^{-1}$). 

To separate the pure activity contribution from the observed line profiles, we apply the spectral subtraction technique with the STARMOD program \citep{Barden1985, Montes1997, Montes2000}. More details on the procedure can be found in Appendix~\ref{seca}.
%-------------------------------------------------------------------------------------------------------------------------%
\begin{figure*}
\centering
\includegraphics[width=15.00cm,height=11.25cm]{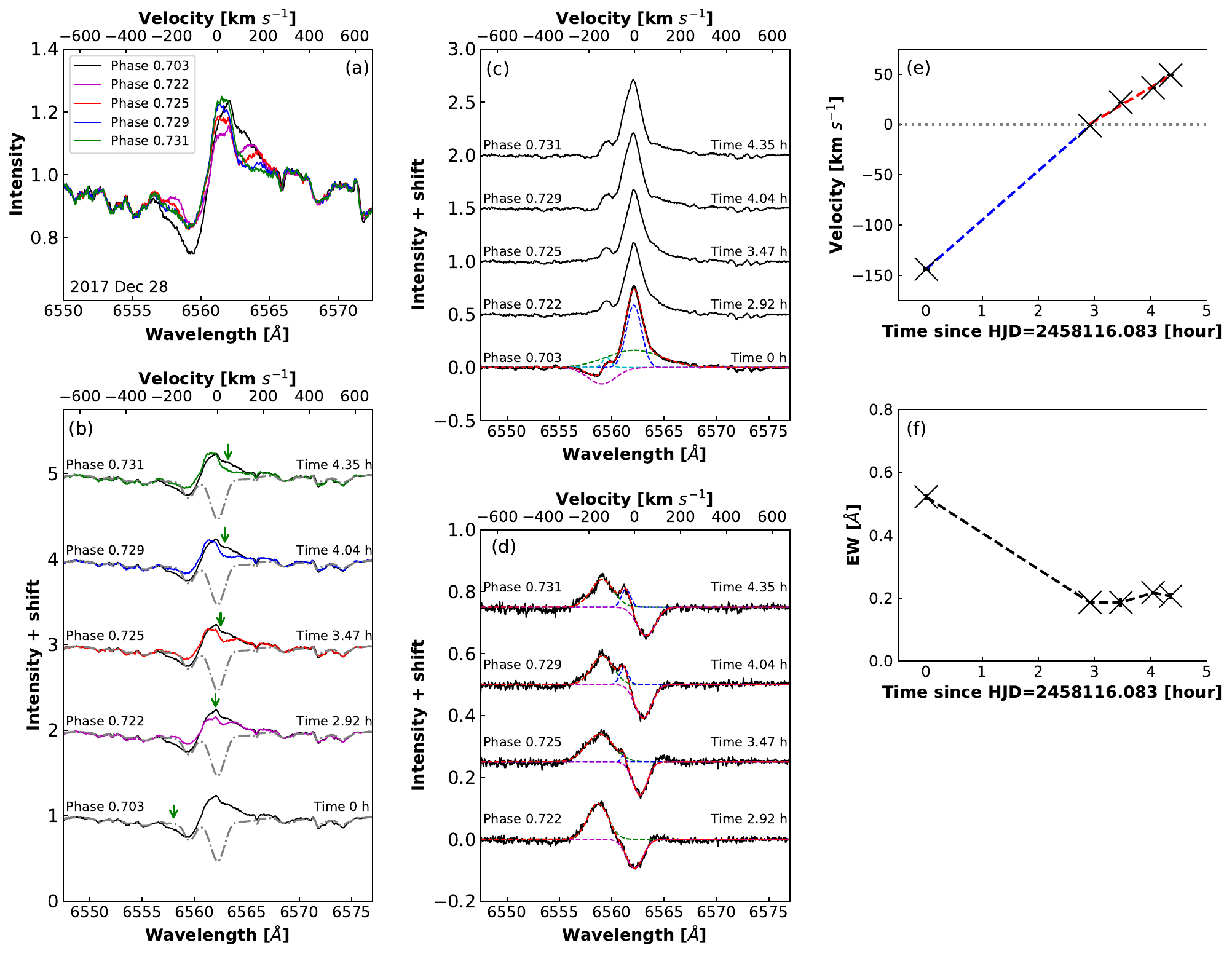}
\caption{H$_{\alpha}$ spectra on 2017 December~28 and Doppler-shifted absorption features in the profiles. (a) Observed H$_{\alpha}$ spectra overlaid for comparison. (b) Observed (solid lines) and synthesized (dash-dotted gray lines) H$_{\alpha}$ spectra. For the latter four observed spectra (colored solid lines), the first observed spectrum is superimposed for comparison. Arrows indicate the absorption features appeared in the H$_{\alpha}$ profiles. (c) Subtracted H$_{\alpha}$ profiles and Gaussian fitting (dashed lines) for the subtraction at phase 0.703. (d) Differences between the first observed H$_{\alpha}$ spectrum and the subsequent four observed spectra (black solid lines), together with the Gaussian fittings (dashed lines). Phases and flare times are marked in above panels, and all spectra are corrected to the rest velocity frame of the K0~IV primary star of UX~Ari. (e) Temporal variation of the bulk velocities of the absorption feature. The blue and red dashed lines represent the linear fittings for the blueshifted velocities and the redshifted ones, respectively. The gray dotted line indicates zero-point of the velocity. (f) Temporal variation of the EWs of the absorption feature.}
\label{Fig2}
\end{figure*}
%-------------------------------------------------------------------------------------------------------------------------%
%%%%%%%%%%%%%%%%%%%%%%%%%%%%%%%%%%%%%%%%%%%%%%%%%%
\section{Analysis and results}\label{sec3}
%-------------------------------------------------------------------------------------------------------------------------%
\subsection{A huge optical flare event}
A long-duration optical flare was hunted in the observations from 2017 December~28 (HJD~=~2458116) to 2018 January~3 (HJD~=~2458122), during which the H$_{\alpha}$ and other chromospheric lines show stronger emission. In particular, the $\mbox{He~{\sc i}}$~D$_{3}$ line, which is typically used as a flare indicator due to its high excitation potential \citep[e.g.][]{Zirin1988, Gu2002, Cao2019, Cao2024}, exhibits obvious emission feature (see Fig.~\ref{Figa1}). Fig.~\ref{Fig1}~(a) shows the EWs of the STARMOD-subtracted H$_{\alpha}$ spectra plotted against HJD. It should be noted that the H$_{\alpha}$ subtractions contain both intrinsic chromospheric emission and flare radiation. Therefore, we remove the intrinsic chromospheric emission from the subtractions to obtain the pure flare radiation. The intrinsic chromospheric emission intensity is obtained by averaging the EWs of two subtracted H$_{\alpha}$ spectra, respectively observed on 2017 December~8 (HJD~=~2458096) and 9 (HJD~=~2458097). And these two spectra were obtained at comparable phases to the spectra on 2017 December~28, with no flare radiation shown in the spectra (hereinafter referred to as pre-flare reference spectra, see Fig.~\ref{Figb1}). The EWs subtracted the intrinsic chromospheric emission are shown in Fig.~\ref{Fig1}~(b), in which we set the observing time (HJD~=~2458116.083) as zero point and then derive the flare time (see the top axis). It can be found that the flare reached its maximum intensity at about 75~hours and the event lasted for a duration of at least 150~hours.

Since the K0~IV primary star is very active in UX~Ari and the subtracted emission is associated with this star during the flare event (see Fig.~\ref{Figa1}), it can be concluded that the optical flare occurred on this star. The flare energy released in the H$_{\alpha}$ line is estimated to be about $1.47~\times~10^{37}$~erg (see the calculation procedure in Appendix~\ref{secc}). Moreover, based on the relationship between the H$_{\alpha}$ and bolometric white-light flare energy from Equation~(2) of \citet{Namekata2024}, the bolometric white-light flare radiation energy is derived to be $1.03~\times~10^{39}$~erg, which suggests that this optical flare is comparable to the most intense stellar superflares.
%-------------------------------------------------------------------------------------------------------------------------%
\subsection{H$_{\alpha}$ line profiles during the flare event} 
\subsubsection{Doppler-shifted absorption feature} 
As shown in Fig.~\ref{Fig2}~(a), five spectra were observed on 2017 November~28, which were during the initial phase of the flare event. It is evident that the H$_{\alpha}$ spectra displayed variable profiles. The first spectrum showed a pronounced absorption in the blue wing, while both the central region and red wing exhibited more intense emission. Furthermore, as illustrated in Fig.~\ref{Fig2}~(b), when compared with the synthesized spectrum, an excess absorption feature really emerged in its blue wing. For the pre-flare reference spectra, there were no similar absorption features (see Fig.~\ref{Figb1}), providing a stronger support for the detection of blueshifted excess absorption feature.
 
In comparison, it can be seen from Fig.~\ref{Fig2}~(a) and (b) that the second spectrum did not show similar enhanced absorption feature in the blue wing and failed to exhibit the anticipated stronger emission in the central region of the line profile compared with the first spectrum during the rising phase of the flare. The observed weaker emission cannot be attributed to a weakening of the flare intensity since other chromospheric activity indicators, such as the $\mbox{He~{\sc i}}$~D$_{3}$ and $\mbox{Ca~{\sc ii}}$ $\lambda$8542 lines (see Fig.~\ref{Figd1}), did not show decreased emission at this time. A plausible explanation for this discrepancy is that the excess absorption feature observed in the blue wing of the first spectrum underwent a wavelength shift in the second spectrum. This shift resulted in localized absorption in the line profile, leading to a weaker central emission and producing a redshifted fake excess emission at the velocity of about 150~km~s$^{-1}$. In the third spectrum, the central emission exhibited an increased intensity relative to the second spectrum, accompanied by a more redshifted fake emission feature in the red wing. This could be attributed to the further redshift of the excess absorption feature, which not only resulted in a more redshifted emission feature but also promoted the recovery and enhancement of the central emission. Furthermore, the variable profiles observed in the final two spectra further supported the detection of a continuous redshift of the excess absorption feature. In summary, the variable H$_{\alpha}$ profiles doubtlessly revealed a rapidly Doppler-shifted excess absorption feature moving from the blue wing to the red one.

With the H$_{\alpha}$ subtraction obtained at phase 0.703, illustrated in Fig.~\ref{Fig2}~(c), we employ the Gaussian fitting method to model the profile and a Gaussian absorption profile is used to represent the excess absorption feature. For the other four subtracted H$_{\alpha}$ spectra, because the absorption feature is completely mixed with the subtracted emission, it is difficult to apply the same modelling method. Therefore, we use the first spectrum as a reference and individually subtract each of the last four observed H$_{\alpha}$ spectra from it to extract the absorption feature. The differences are shown in Fig.~\ref{Fig2}~(d) and modelled using the Gaussian fitting method. As previous description, the variations observed in the line profiles of the last four spectra are mainly attributed to the localized absorption feature that had shifted to the central regions and red wings of these spectral profiles. Moreover, this excess absorption appeared in the blue wing of the first profile. Therefore, it is appropriate to use the first profile as a reference, particularly its emission region, to isolate the absorption features presented in the last four profiles. Although this approach may not accurately determine the intensity of the absorption feature, it can provide information regarding its velocity variation. In addition, the presence of excess absorption in the blue wing of the first profile results in a localized stronger emission feature in each difference spectrum. In contrast, weaker emission is also obtained in the last three difference spectra, resulting from changes in the line profiles, particularly their central regions evolved into stronger emission. The fitting results indicate that the absorption feature firstly appeared with a velocity of -143.9 km~s$^{-1}$ at phase 0.703, then decelerated to -1.1 km~s$^{-1}$ at phase 0.722, next turned into a redshifted absorption, and finally moved to 49.2 km~s$^{-1}$ at phase 0.731 over the course of about 4.4 hours. The velocities and EWs of the absorption features are plotted against flare time in Fig.~\ref{Fig2}~(e) and (f), respectively.
%-------------------------------------------------------------------------------------------------------------------------%
\subsubsection{Far-redshifted excess emission features}
For the observations on 2017 December~31, 2018 January~1 and 3, there are far-redshifted emission components in the fitted H$_{\alpha}$ subtractions, indicating the presence of other forms of plasma motion during the flare (see more details in Appendix~\ref{sece}).
%------------------------------------------------------------------------------------------------------------%
\begin{figure*}
\centering
\includegraphics[width=6cm,height=4.25cm]{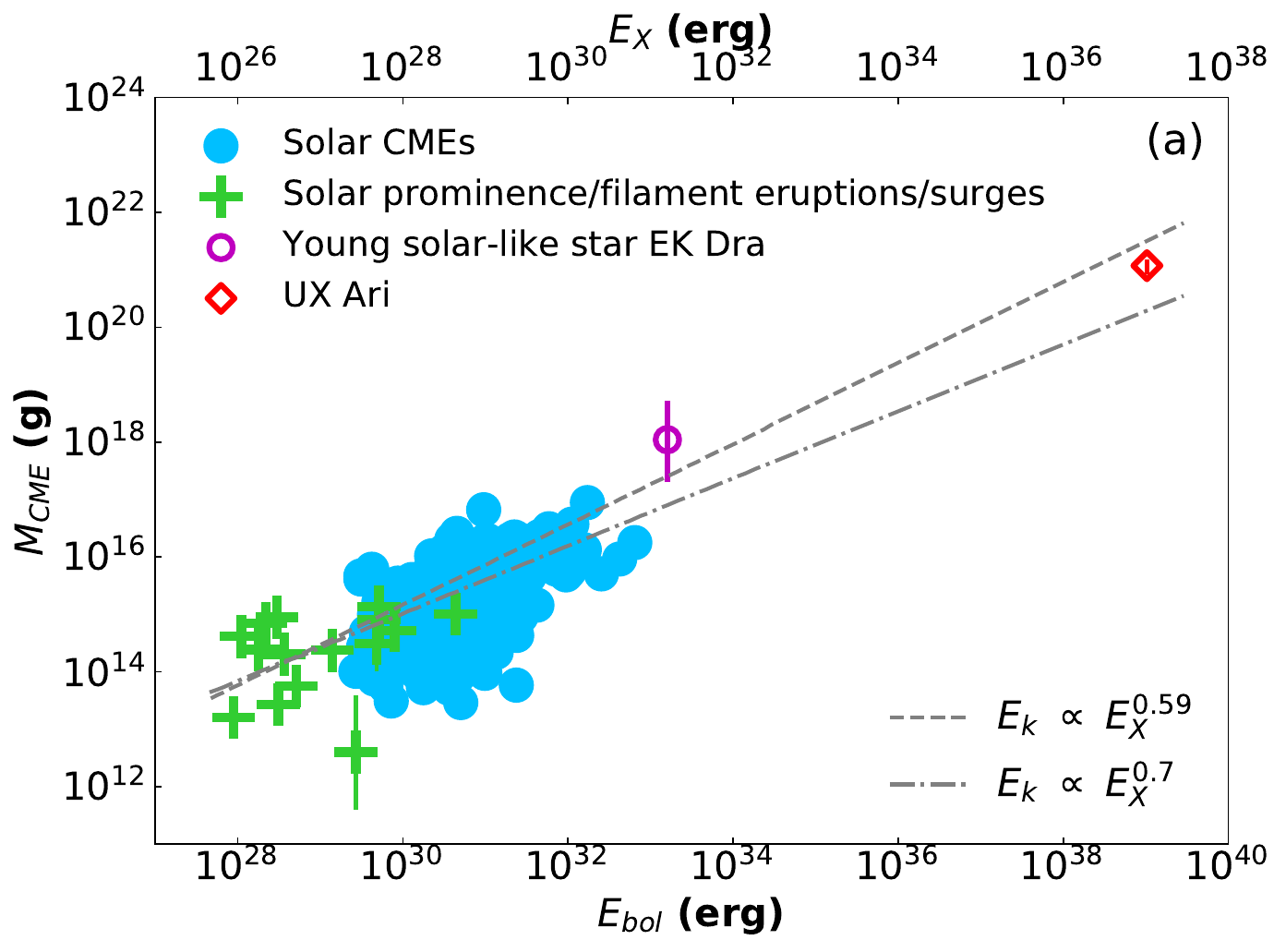}
\includegraphics[width=6cm,height=4.25cm]{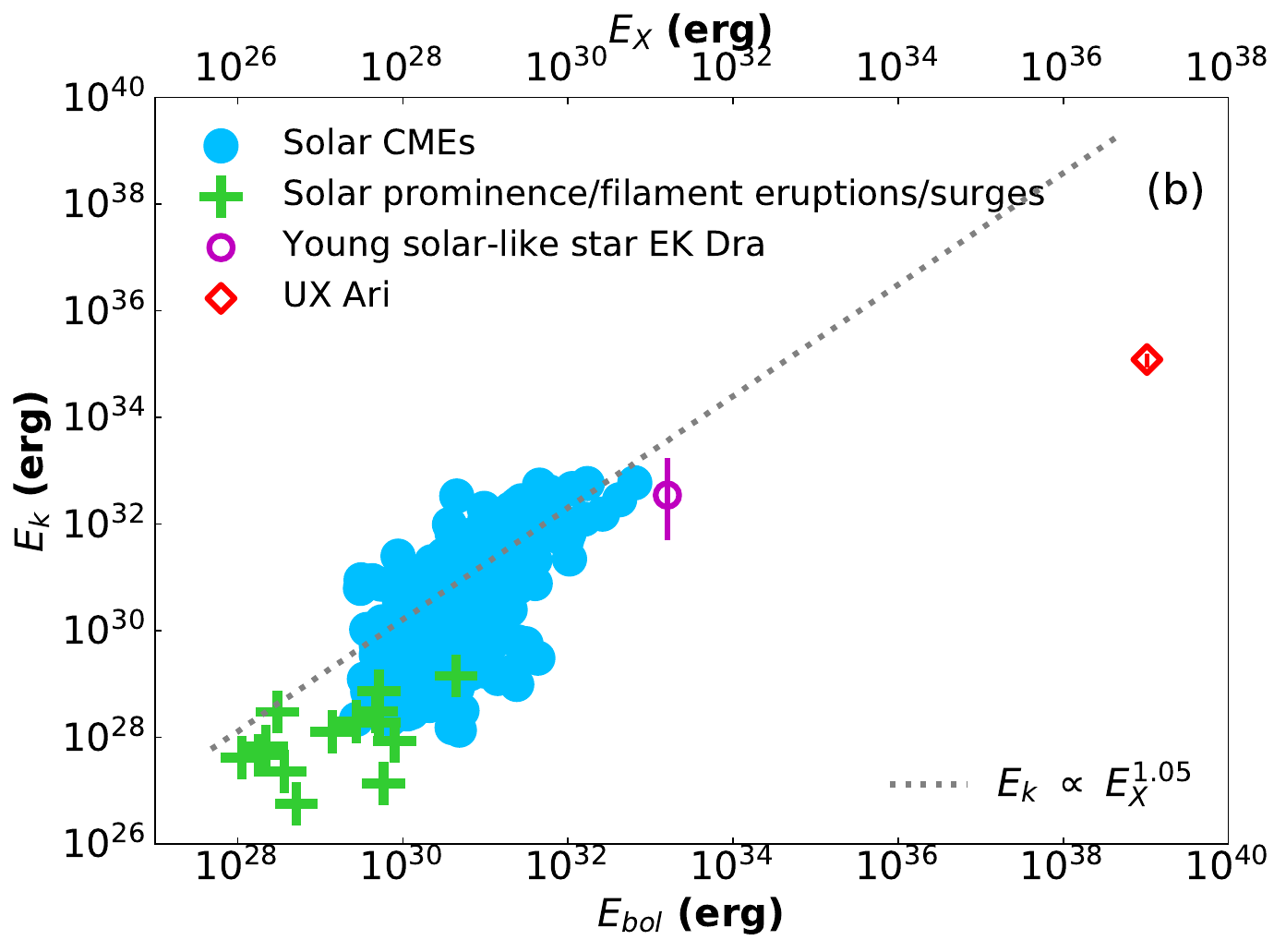}
\includegraphics[width=6cm,height=4.0cm]{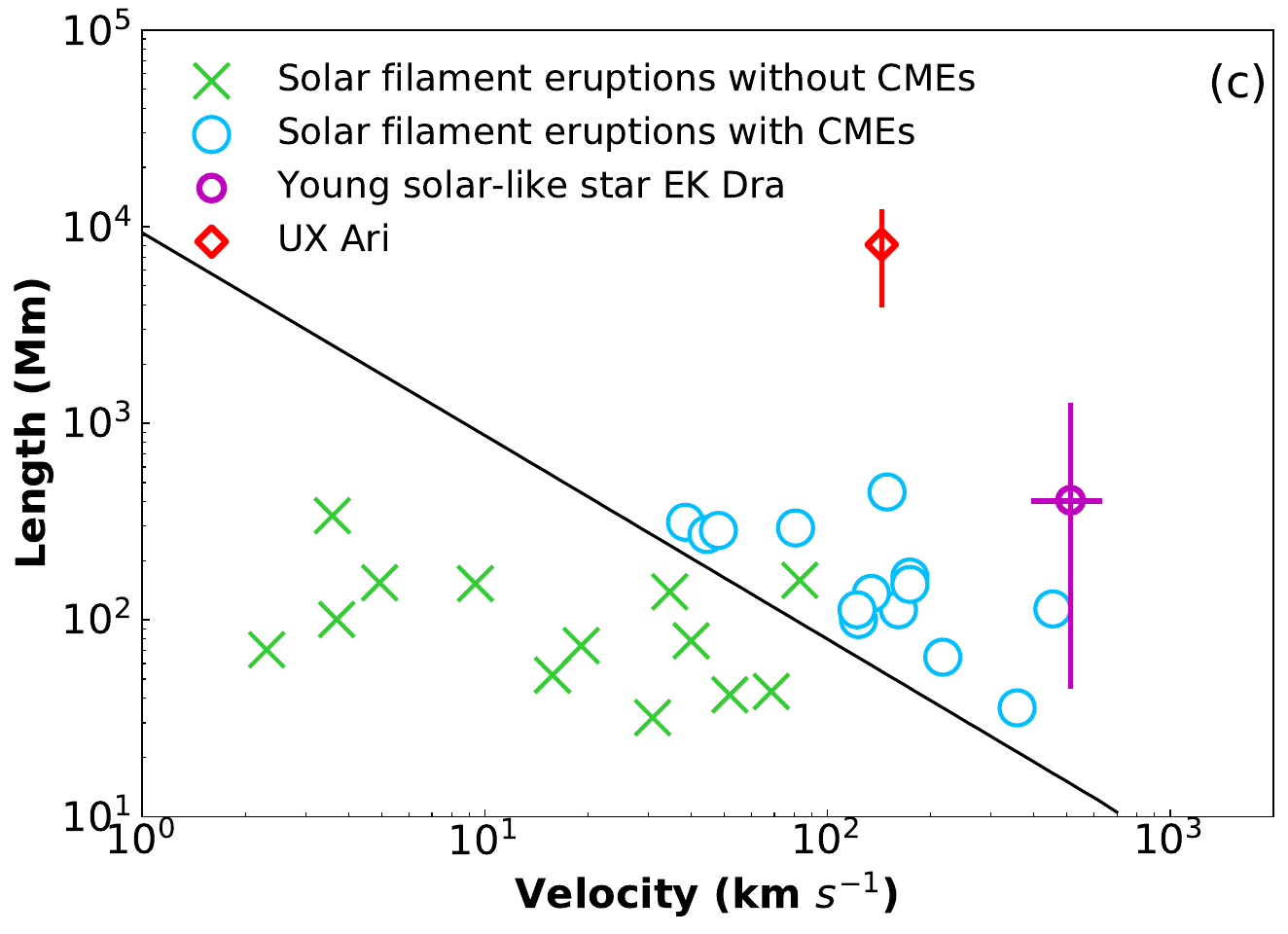}
\caption{Mass $M_{CME}$ (a) and kinetic energy $E_{k}$ (b) are plotted as functions of flaring bolometric energy (bottom horizontal axis) and X-ray energy emitted in the GOES band (top axis) for solar and stellar filament/prominence eruptions and CMEs. The solar CME data (filled sky-blue circles) are derived from \citet{Yashiro2009}. The solar prominence/filament eruptions/surges (green plus) are obtained from \citet{Namekata2024}, originally taken from \citet{Namekata2022} and \citet{Kotani2023}. The dashed and dotted lines are trend fits for the solar CMEs expressed as $M_{CME}$~$\propto$~$E_{X}^{0.59}$ and $E_{k}$~$\propto$~$E^{1.05}$ \citep{Drake2013} and $M_{CME}$~$\propto$~$E_{X}^{0.7}$ \citep{Aarnio2012}. (c) Comparison between the velocity and length scales of the filament/prominence eruptions on the Sun and UX~Ari. The solid black line indicates a threshold \citep[see][]{Seki2021}.}
\label{Fig3}
\end{figure*}
%------------------------------------------------------------------------------------------------------------%
%%%%%%%%%%%%%%%%%%%%%%%%%%%%%%%%%%%%%%%%%%%%%%%%%%%%%%%%
\section{Discussion}\label{sec4}
\subsection{A filament eruption during the initial phase of the flare}
\subsubsection{Rapidly Doppler-shifted absorption feature indicating a failed filament eruption}
The blueshifted absorption feature with a bulk velocity of -143.9~km~s$^{-1}$ implies the existence of low-temperature and high-density neutral plasma above the stellar disk, which is moving toward the observer at a higher velocity. A filament eruption is a reliable explanation. A similar absorption feature resulting from a filament eruption connected to a superflare had also been found by \citet{Namekata2022} on a young solar-type star EK~Dra, and further verified by analyzing a solar filament eruption with the Sun-as-a-star method. For our situation, the bulk velocity of the blueshifted absorption is much smaller than the escape velocity of the K0~IV primary star of UX~Ari. The escape velocity is estimated to be 304~km~s$^{-1}$ from $v_{esp}~=~630(\frac{M_{\star}}{M_{\sun}})^{1/2}(\frac{R_{\star}}{R_{\sun}})^{-1/2}$~km~s$^{-1}$ with the mass $M_{\star}~=~1.3~M_{\sun}$ and the radius $R_{\star}~=~5.6~R_{\sun}$ \citep{Hummel2017}. Moreover, even when a projection angle of 45$^{\circ}$ is assumed, the velocity ($\sim$-204~km~s$^{-1}$) is still far below the escape velocity, which suggests that the erupted filament was unable to escape from the star and also indicates that the filament would fall back to the star.

The blueshifted absorption feature decelerated to -1.1~km~s$^{-1}$ at phase~0.722, implying that the filament nearly erupted to its highest point at this phase. The average deceleration rate from phase 0.703 to 0.722 is approximately 13.6~m~s$^{-2}$, which exceeds the deceleration of about 11.5~m~s$^{-2}$ due to the surface gravity of the K0~IV primary star (log~$g$~=~3.06~cm~s$^{-2}$, \citealt{Hummel2017}). This suggests that an additional force was influencing the filament eruption in addition to the gravity. The strong magnetic field on UX~Ari can account for this additional force because the erupted filament might be suppressed by the strong overlying magnetic field of the active star \citep[see][]{Alvarado2018}. The erupted filament could reach a minimum height of about $7.5~\times~10^{5}$~km ($\sim$~0.19 stellar radii of the K0~IV star) based on the deceleration rate.

After phase 0.722, the absorption feature became accelerated redshift (see Fig.~\ref{Fig2}~(e)), suggesting that the erupted filament material was falling back toward the star and it is also consistent with the previous point of view that the filament would eventually fall back to the star. The average acceleration rate during the falling phase was about 9.4~m~s$^{-2}$, which is slightly less than the value of the surface gravity. This indicates that the angle between the filament's falling direction and the line of sight is relatively insignificant. Therefore, a projection angle of about 35$^{\circ}$ is derived from $i~=~\arccos(9.4/11.5)$ when the filament is subjected to the gravity and falling. This also implies that the previous assumption of 45$^{\circ}$ about the projection angle is reasonable.

To sum up, a filament eruption occurred during our observations. The filament showed a gradual decrease in the eruption velocity, then reached its highest point, and finally fell back toward the star. Therefore, this fact is indicative of a failed filament eruption. Although the filament eruption occurred during the initial phase of the flare, the spectral line had already exhibited an increase in intensity prior to this event. Therefore, it is challenging to definitively conclude that the filament eruption was responsible for triggering the flare. However, it remains possible that the filament had already begun to become unstable before the flare.
%-------------------------------------------------------------------------------------------------------------------------%
\subsubsection{Properties of the filament eruption}
The filament mass is derived to be $8.5-15.0~\times~10^{20}$~g (see Appendix~\ref{secf}), which is consistent with the masses of the erupted prominences detected on other active stars \citep[e.g.][]{Inoue2023, Cao2024}. Moreover, the kinetic energy is derived to be $8.8-15.5~\times~10^{34}$~erg. Fig.~\ref{Fig3} (a) and (b) show the mass and kinetic energy as functions of the flare energy emitted in the GOES X-ray and bolometric flare energy in comparison with solar filament/prominence eruptions and CMEs. The prescription for conversion of the flare energy can be found in \citet{Cao2024}. It can be seen that the erupted filament mass is basically consistent with the prediction of the solar flare-CME relation, indicating that the filament eruption has a similar underlying mechanism with solar eruptions. This result is in good agreement with properties of prominence eruptions found on other active stars \citep[e.g.][]{Moschou2019, Namekata2022, Inoue2023, Cao2024, Namekata2024}. Moreover, similar to findings on other active stars, the kinetic energy is less than that expected from solar CMEs. One potential explanation, as discussed in the previous section, is that the overlying magnetic field on UX~Ari could suppress the filament eruption. And this could result in the low kinetic energy. Another probable reason is that inner filament/prominence eruptions typically exhibit lower velocities than outer hotter CME plasmas, as observed on the Sun. However, it is important to note that the kinetic energy derived by us is 4–5 orders of magnitude smaller than that predicted by the solar observations and is also significantly lower than that observed on other stars. Typically, the kinetic energies are about 2~orders of magnitude lower than that predicted from the solar extrapolation on other active stars \citep{Moschou2019}, and there are also some studies that had reported kinetic energies consistent with the solar flare-CME relation \citep[e.g.][]{Inoue2023}. For the significantly lower kinetic energy, there might be other contributing factors, such as the possible occurrence of a sympathetic eruption (see the discussion in Sec.~4.2).

Despite erupted filaments/prominences finally return to the Sun, CMEs can still occur. \citet{Namekata2022} suggested that a CME happened on EK~Dra through comparing the length scale and velocity of the filament eruption with ones of solar filament/prominence eruptions. The same method is adopted here, as shown in Fig.~\ref{Fig3}~(c), in which the length scale and velocity of the filament eruption on UX~Ari is compared with those of solar eruptions. It can be seen that the filament eruption of UX~Ari is above the threshold, which can distinguish whether a filament eruption is associated with a CME or not. This suggests that the filament eruption on UX~Ari may be associated with a CME. Nevertheless, this conclusion is still questionable since the threshold is derived from solar observations. EK~Dra is a Sun-like star, so the threshold may be potentially applicable. The magnetic field strength and topology of UX~Ari, as well as its atmospheric parameters, are different from those of the Sun and the scale of the filament eruption is considerably larger than that observed on the Sun (see Fig.~\ref{Fig3}). These discrepancies may imply that the threshold maybe is not applicable here. Therefore, it is not the time to draw any conclusion for this point, we leave it as an open question to future follow-up studies.
%-------------------------------------------------------------------------------------------------------------------------%
\subsection{Potential flaring location and its possible spatial relationship with the filament eruption}
The broad emission component in the subtracted flaring spectra is dominant, and its Doppler-shifted velocity ranges approximately from -19 to 20 km s$^{-1}$ over phases from 0.012 to 0.649, spanning more than half of the period. This suggests that the Doppler shift of the broad emission component is likely attributable to the stellar rotation. A cosine function is utilized to model the velocity variation, resulting in a maximum blueshifted velocity of about -25 km s$^{-1}$ on HJD~=~2458118.285 (at phase~0.045). This occurs when the flare was located at the stellar limb that is moving towards us. Moreover, the most blueshifted velocity and the $v \sin i$ value of 39 km s$^{-1}$ for the K0~IV primary star imply that the flare occurred at a latitude of about 50$^{\circ}$ ($\sim$~arccos(25/39)). The inclination of UX~Ari is about 59.2$^{\circ}$ \citep{Duemmler2001}, which implies that the regions of the photosphere located above latitude of 59.2$^{\circ}$ are consistently visible. Because the flare ribbon occurred in the chromosphere which is higher than the photosphere, it is possible that the flare happened at a latitude of about 50$^{\circ}$ still is always visible.

The filament eruption took place on the hemisphere oriented towards us on 2017 December~28, whereas the flare was located at the opposing hemisphere (see Fig.~\ref{Fige2} (a)). Therefore, if the filament eruption originated from the flare with a highly inclined trajectory relative to the vertical direction, the inclination angle would exceed 90$^{\circ}$. Although solar filament eruptions can occasionally occurred at significantly inclined angles \citep{Seki2019}, no instances of such extreme inclination angles have been found in solar observations. Alternatively, another possible scenario is that the filament eruption and the flare, despite occurring nearly at the same time, were located at different regions. In solar observations, some successive eruption events occurred at different active regions may be physically interconnected. This phenomenon is referred to as sympathetic eruptions, wherein the occurrence of one eruption event triggers another at a different location \citep{Wheatland1998}. Therefore, for our situation,  the relationship between the filament eruption and the flare may be indicative of a sympathetic eruption.
%%%%%%%%%%%%%%%%%%%%%%%%%%%%%%%%%%%%%%%%%%%%%%%%%%%%%%%%
\section{Summary and conclusions}\label{sec5}
A huge optical flare event was detected on the active RS~CVn-type star UX~Ari, which lasted for a duration of at least 150 hours. The flare energy released in the H$_{\alpha}$ line is estimated to be about $1.47 \times 10^{37}$~erg, and the bolometric white-light flare energy is converted to be $1.03 \times 10^{39}$~erg. A rapidly Doppler-shifted absorption feature moving from the blue wing of the H$_{\alpha}$ line profile to the red one was identified during the initial phase of the flare, which can be ascribed to a filament eruption. The filament finally fell back toward the star and therefore it is a possible failed filament eruption. The erupted filament mass of $8.5-15.0~\times~10^{20}$~g and the kinetic energy of $8.8-15.5~\times~10^{34}$~erg are derived. The mass value is in good agreement with the solar flare-CME relation. However, the value of kinetic energy is significantly lower than that predicted by this relation. Furthermore, the relationship between the filament eruption and the flare is likely characterized by a sympathetic eruption.
%%%%%%%%%%%%%%%%%%%%%%%%%%%%%%%%%%%%%%%%%%%%%%%%%%%%%%%%
\begin{acknowledgements}
The authors thank the anonymous referee for the careful review and helpful suggestions, which lead to a significant improvement in our manuscript. The authors would like to express their gratitude to the staff of the Xinglong 2.16m telescope for their assistance and support. This work is partially supported by the Open Project Program of the Key Laboratory of Optical Astronomy, National Astronomical Observatories, Chinese Academy of Sciences. The present study is also financially supported by the National Natural Science Foundation of China (NSFC) under grant Nos. 10373023, 10773027, 11333006, and U1531121, the Yunnan Fundamental Research Projects (grant Nos.~202201AT070186 and 202305AS350009), the Yunnan Revitalization Talent Support Program (Young Talent Project), and International Centre of Supernovae, Yunnan Key Laboratory (No.~202302AN360001). The authors also acknowledge the science research grant from the China Manned Space Project.
\end{acknowledgements}

\bibliographystyle{aa}
\bibliography{aa} 
%%%%%%%%%%%%%%%%%%%%%%%%%%%%%%%%%%%%%%%%%%%%%%%%%%%%%%%%
%%appendix
\begin{appendix} %First  appendix
\nolinenumbers
\section{Observing log of UX~Ari and spectral subtraction}\label{seca}
Table~\ref{taba1} presents the observing log of UX~Ari, which includes the observing date, exposure time (Exp. time), heliocentric Julian date (HJD), and orbital phase calculated with the ephemeris HJD = 2, 456, 238.134 + $6.437888P$ from \citet{Hummel2017}. The zero phase corresponds to the time when the G5~V secondary component is at inferior conjunction in front of the K0~IV primary component. The long-duration optical flare was observed from 2017 December~28 to 2018 January~3, while the observations on 2017 December~8 and 9 were used to obtain the intrinsic chromospheric emission during the flare.
%------------------------------------------------------------------------------------------------------------%
\begin{table}
\centering
\caption{Observing information for UX~Ari.}
\tabcolsep 0.3cm
\label{taba1}
\begin{tabular}{lccccr}
\hline
\multicolumn{1}{l}{Date} &
\multicolumn{1}{r}{Exp.time}&
\multicolumn{1}{c}{HJD} &
\multicolumn{1}{c}{Phase}\\
   &(s)&(2,450,000+)&\\
\hline
2017/12/08 & 900   & 8096.276 & 0.626\\
2017/12/09 & 900   & 8097.181 & 0.767\\
2017/12/28 & 1200 & 8116.083 & 0.703\\
2017/12/28 & 900   & 8116.205 & 0.722\\
2017/12/28 & 900   & 8116.228 & 0.725\\
2017/12/28 & 900   & 8116.252 & 0.729\\
2017/12/28 & 900   & 8116.265 & 0.731\\
2017/12/30 & 900   & 8118.072 & 0.012\\
2017/12/30 & 1200 & 8118.111 & 0.018\\
2017/12/31 & 1200 & 8119.075 & 0.167\\
2017/12/31 & 1200 & 8119.154 & 0.180\\
2017/12/31 & 1200 & 8119.170 & 0.182\\
2017/12/31 & 1200 & 8119.202 & 0.187\\
2018/01/01 & 1200 & 8120.072 & 0.322\\
2018/01/01 & 900   & 8120.132 & 0.332\\
2018/01/01 & 900   & 8120.145 & 0.334\\
2018/01/01 & 900   & 8120.171 & 0.338\\
2018/01/01 & 900   & 8120.184 & 0.340\\
2018/01/02 & 3600 & 8121.100 & 0.482\\
2018/01/02 & 3600 & 8121.144 & 0.489\\
2018/01/03 & 3600 & 8122.131 & 0.642\\
2018/01/03 & 3600 & 8122.175 & 0.649\\
\hline
\end{tabular}
\end{table}
 %------------------------------------------------------------------------------------------------------------%
 
\begin{figure}
\centering
\includegraphics[width=8.5cm,height=8.5cm]{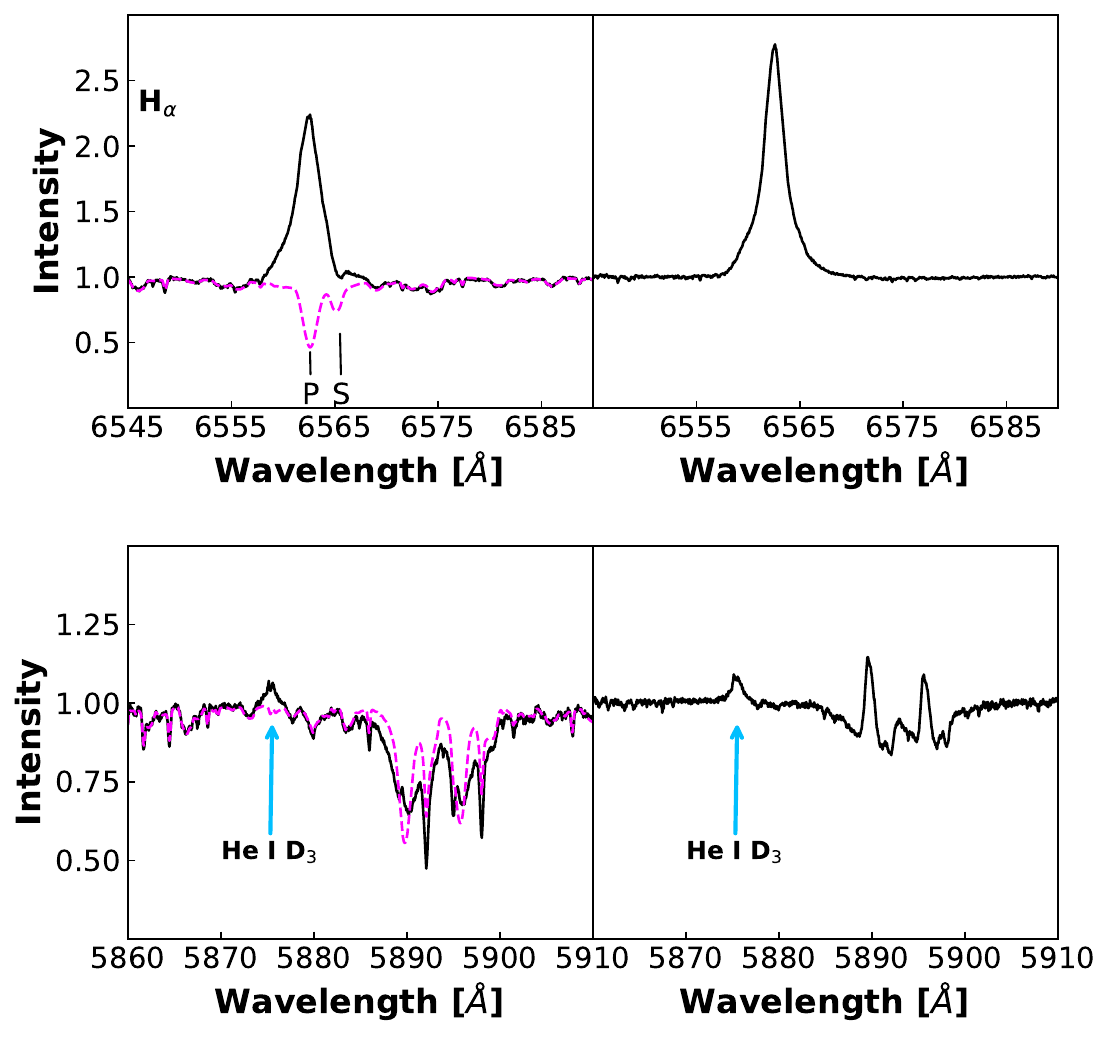}
\caption{Examples of the observed, synthesized, and subtracted spectra for the H$_{\alpha}$ (upper panel) and $\mbox{He~{\sc i}}$ D$_{3}$ line (lower panel) spectral regions in one flaring spectrum obtained at phase 0.322 on 2018 January~1. For the left part of each panel, the black solid line is the observed spectrum, and the magenta dashed line represents the synthesized spectrum. ``P'' and ``S" indicate the primary and secondary components of UX~Ari, respectively. The resulting subtraction spectrum is displayed in the right part. The label identifying each chromospheric line is also marked.}
\label{Figa1}
\end{figure}

Spectral subtraction technique has been widely used for chromospheric activity studies in active binary systems. The STARMOD program can construct a synthesized spectrum through artificially rotationally broadening, intensity weighting, and radial velocity shifting for the spectra of the reference inactive stars with the similar spectral types and luminosity classes as the two components of the binary system. Subsequently, this synthesized spectrum is subtracted from the observed spectrum of the binary system. The synthesized spectrum represents the non-active state of the system, and therefore the subtraction between the observed and synthesized spectra comes from the pure activity. The activity contribution can be obtained by measuring equivalent widths (EWs) of the subtracted line profiles. For our situation, the spectra of two inactive stars HR~3351 (K0~IV) and HR~3309 (G5~V) are used as references for the K0~IV primary and G5~V secondary star of UX~Ari, respectively. The $vsini$ values of 39~km~s$^{-1}$ for the primary component and 7.5~km~s$^{-1}$ for the secondary component are used, and the intensity weight ratios of two components are 0.74/0.26 for the H$_{\alpha}$ spectral region and 0.69/0.31 for the $\mbox{He~{\sc i}}$ D$_{3}$ spectral region \citep{Cao2017}. As examples, we show the chromospheric lines of one spectrum and illustrate the above described processing in Fig.~\ref{Figa1}.
 %-------------------------------------------------------------------------------------------------------------------------%
%-------------------------------------------------------------------------------------------------------------------------%
%%%%%%%%%%%%%%%%%%%%%%%%%%%%%%%%%%%%%%%%%%%%%%%%%%%%%%%%
\section{Pre-flare reference spectra}\label{secb}
 %-------------------------------------------------------------------------------------------------------------------------%
 \begin{figure}
\includegraphics[width=8.5cm,height=8.5cm]{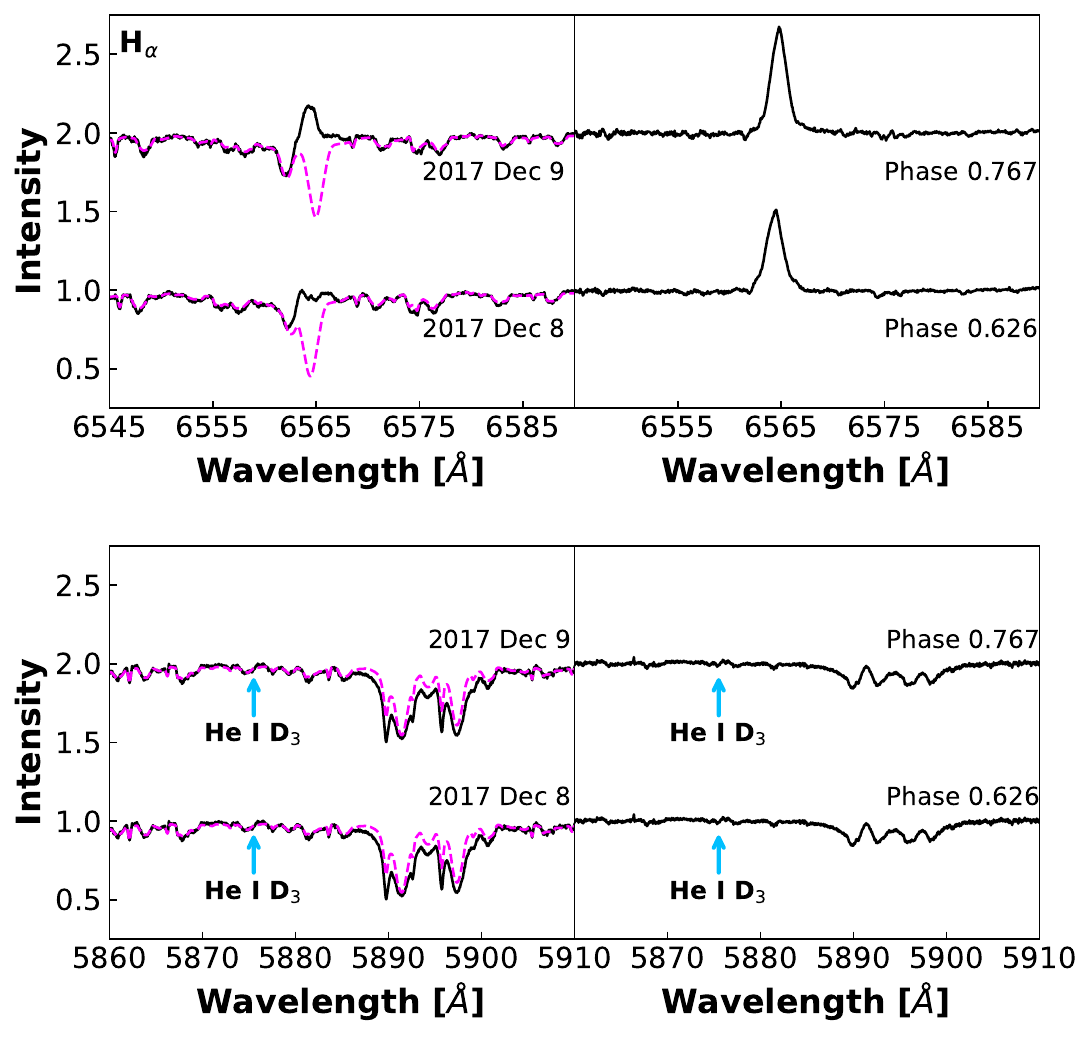}
\includegraphics[width=8.25cm,height=6.25cm]{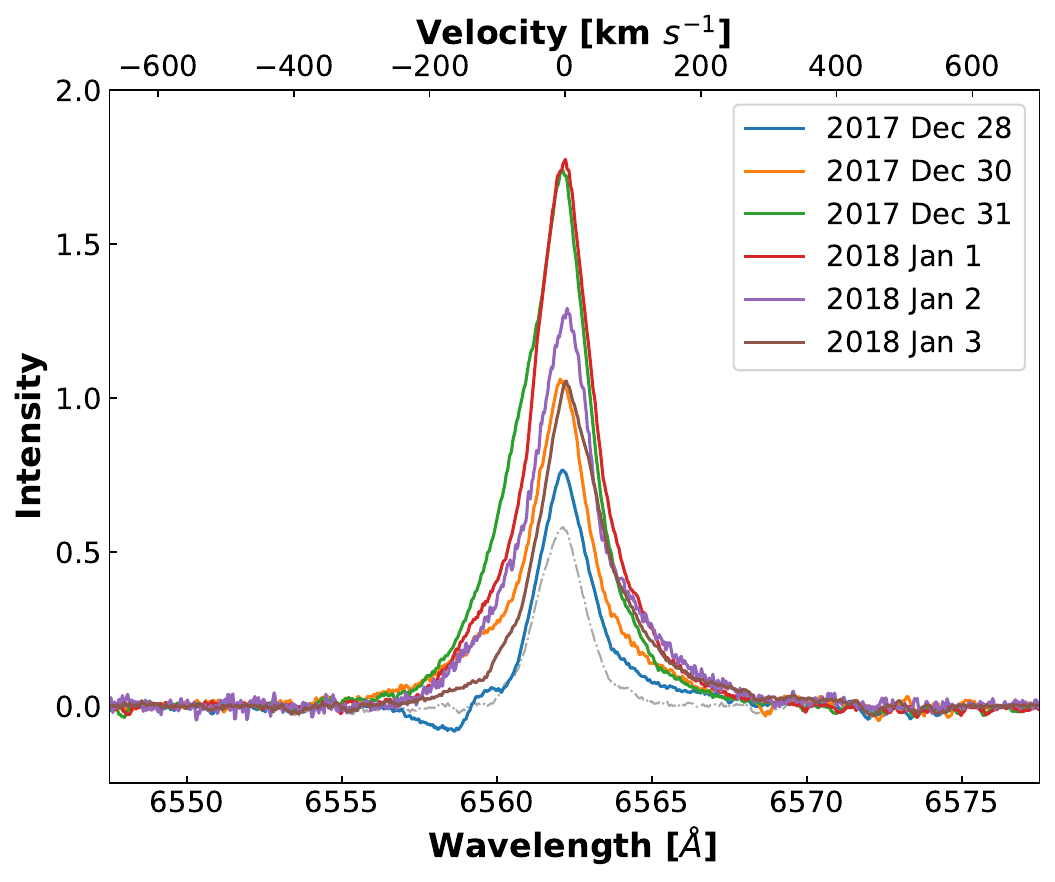}
\caption{Upper and middle panels are the same as Fig.~\ref{Figa1}, but for the spectra observed at phase 0.626 on 2017 December 8 and at phase 0.767 on 2017 December 9. The mean subtracted H$_{\alpha}$ spectrum (gray dash-dotted line) is obtained for the two subtractions and plotted together with the subtracted flaring spectra in the bottom panel. For each observing night, one subtracted flaring spectrum is presented as an example. And all subtracted spectra are corrected to the rest velocity frame of the K0~IV primary star of UX~Ari.}
\label{Figb1}
\end{figure}
%-------------------------------------------------------------------------------------------------------------------------%
The pre-flare reference spectra observed on 2017 December 8 (HJD~=~2458096) and 9 (HJD~=~2458097) are shown in Fig.~\ref{Figb1}, in comparison with the synthesized spectra. Moreover, for two H$_{\alpha}$ subtractions of the pre-flare reference spectra after removing the synthesized spectra, we also calculate a mean spectrum and compared it with the subtracted flaring spectra in the figure.

When analyzing the spectral line profiles, we do not subtract the per-flare reference spectra from the observed flaring spectra. The reason is that UX~Ari is a highly active RS CVn-type star, so that its chromospheric line profile can vary significantly among different nights. A gap of some days existed between the observation of the pre-flare reference spectra and that of the flaring spectra. Consequently, we only opted to use the pre-flare reference spectra to determine the intrinsic chromospheric emission intensity during the flare.
%-------------------------------------------------------------------------------------------------------------------------%
%%%%%%%%%%%%%%%%%%%%%%%%%%%%%%%%%%%%%%%%%%%%%%%%%%%%%%%%
\section{Calculation of the flare energy released in the H$_{\alpha}$ line}\label{secc}
 %-------------------------------------------------------------------------------------------------------------------------%
For our situation, a few observations were made per night during the flare and therefore we calculate a mean EW for the flare emission in each observing night. We compute the stellar continuum flux $F_{H{\alpha}}$~(in erg cm$^{-2}$ s$^{-1}$ \AA$^{-1}$) near the H$_{\alpha}$ line region as a function of the color index $B-V$ ($\sim$~0.88 for UX~Ari; \citealt{Aarum2003a}) based on the empirical relationship: 
\begin{eqnarray}
\log{F_{H{\alpha}}}=[7.538-1.081(B-V)]\pm{0.33}\nonumber \\
0.0~\leq~B-V~\leq~1.4
\end{eqnarray}
from \citet{Hall1996}, and then convert the mean EW into an absolute surface flux $F_{S}$~(in erg~cm$^{-2}$~s$^{-1}$). The EW value is corrected to the total continuum before converting to the absolute surface flux. Therefore, the mean flare luminosity per observing night is obtained by using the derived absolute surface flux and the stellar radius $R_{\star}$~=~5.6~$R_{\sun}$ \citep{Hummel2017} of the K0~IV primary star.

To estimate the flare energy, we integrate the mean luminosity by 48~hours for the observation on 2017 December~28 and 24~hours for the other observations, and then add them together. Because there is an observing gap between 2017 December~28 and December~30, the integrating time of 48~hours is used for the observation on December~28. Therefore, the flare energy released in the H$_{\alpha}$ line is estimated to be about $1.47~\times~10^{37}$~erg.
%
%-------------------------------------------------------------------------------------------------------------------------%
%%%%%%%%%%%%%%%%%%%%%%%%%%%%%%%%%%%%%%%%%%%%%%%%%%%%%%%%
\section{Observed $\mbox{He~{\sc i}}$~D$_{3}$ and $\mbox{Ca~{\sc ii}}$ $\lambda$8542 lines on 2017 December~28}\label{secd}
\begin{figure}
\centering
\includegraphics[width=8.5cm,height=10.5cm]{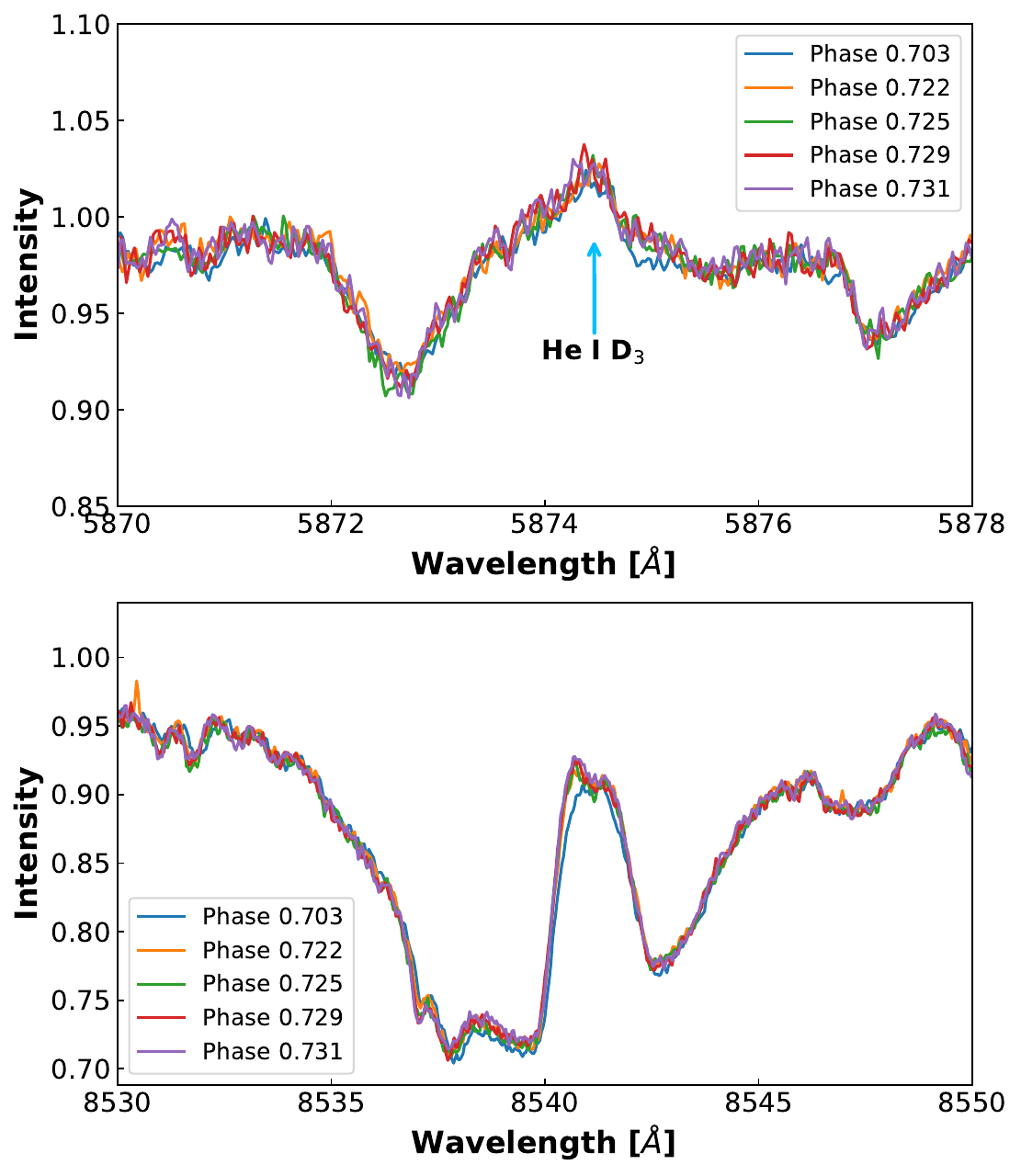}
\caption{Observed $\mbox{He~{\sc i}}$~D$_{3}$ (upper panel) and $\mbox{Ca~{\sc ii}}$ $\lambda$8542 (bottom panel) line profiles on 2017 December~28. The velocities of all spectra are corrected relative to the K0~IV primary star of UX~Ari.}
\label{Figd1}
\end{figure}

Fig.~\ref{Figd1} presents the observed $\mbox{He~{\sc i}}$~D$_{3}$ and $\mbox{Ca~{\sc ii}}$ $\lambda$8542 line profiles on 2017 December~28, demonstrating progressively enhanced emission features consistent with the expected variation during the rising phase of the stellar flare.
%-------------------------------------------------------------------------------------------------------------------------%
%
%-------------------------------------------------------------------------------------------------------------------------%
%%%%%%%%%%%%%%%%%%%%%%%%%%%%%%%%%%%%%%%%%%%%%%%%%%%%%%%%
\section{Far-redshifted emission components}\label{sece}
%-------------------------------------------------------------------------------------------------------------------------%
\begin{figure*}
\centering
\includegraphics[width=8.75cm,height=12cm]{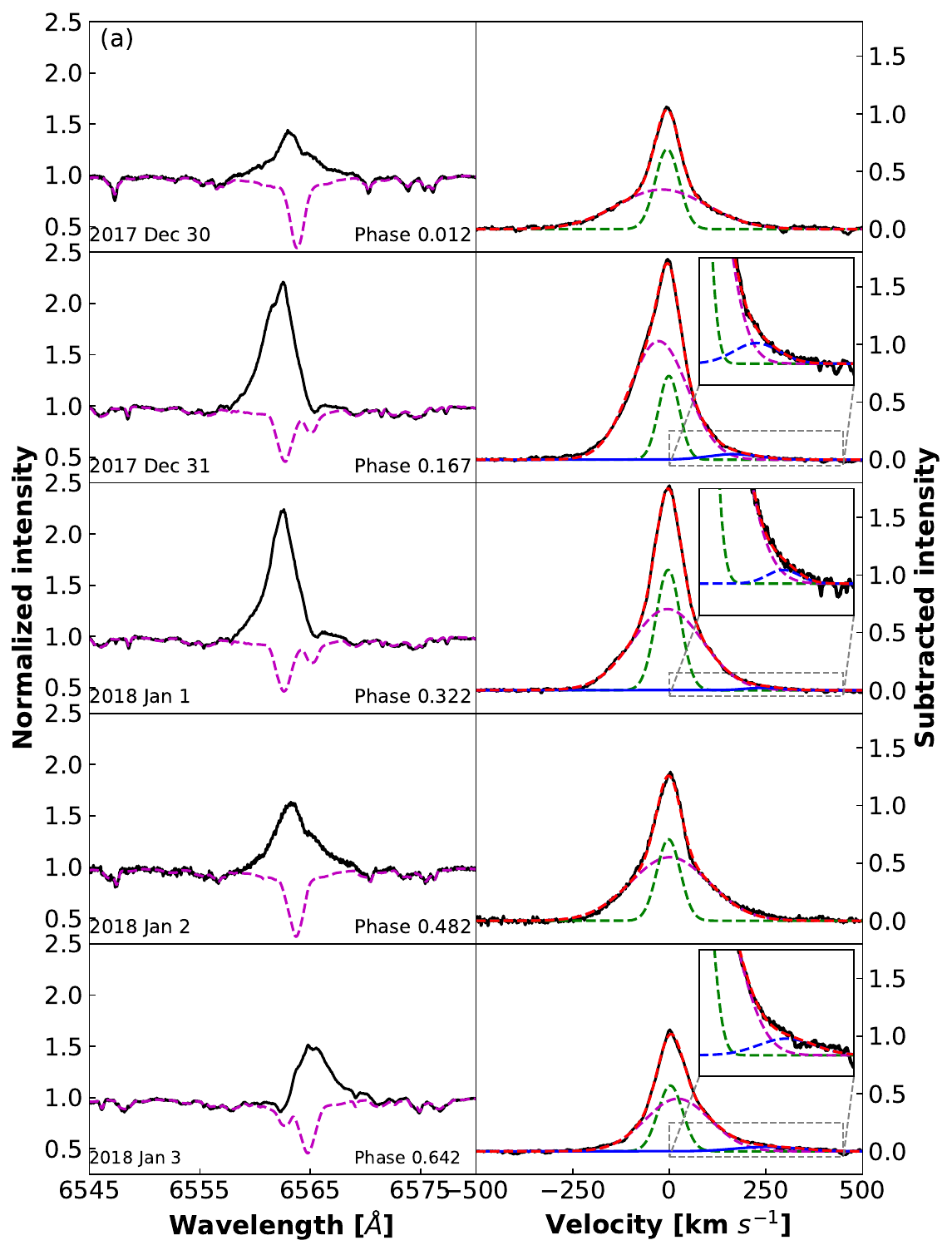}
\includegraphics[width=8.75cm,height=12cm]{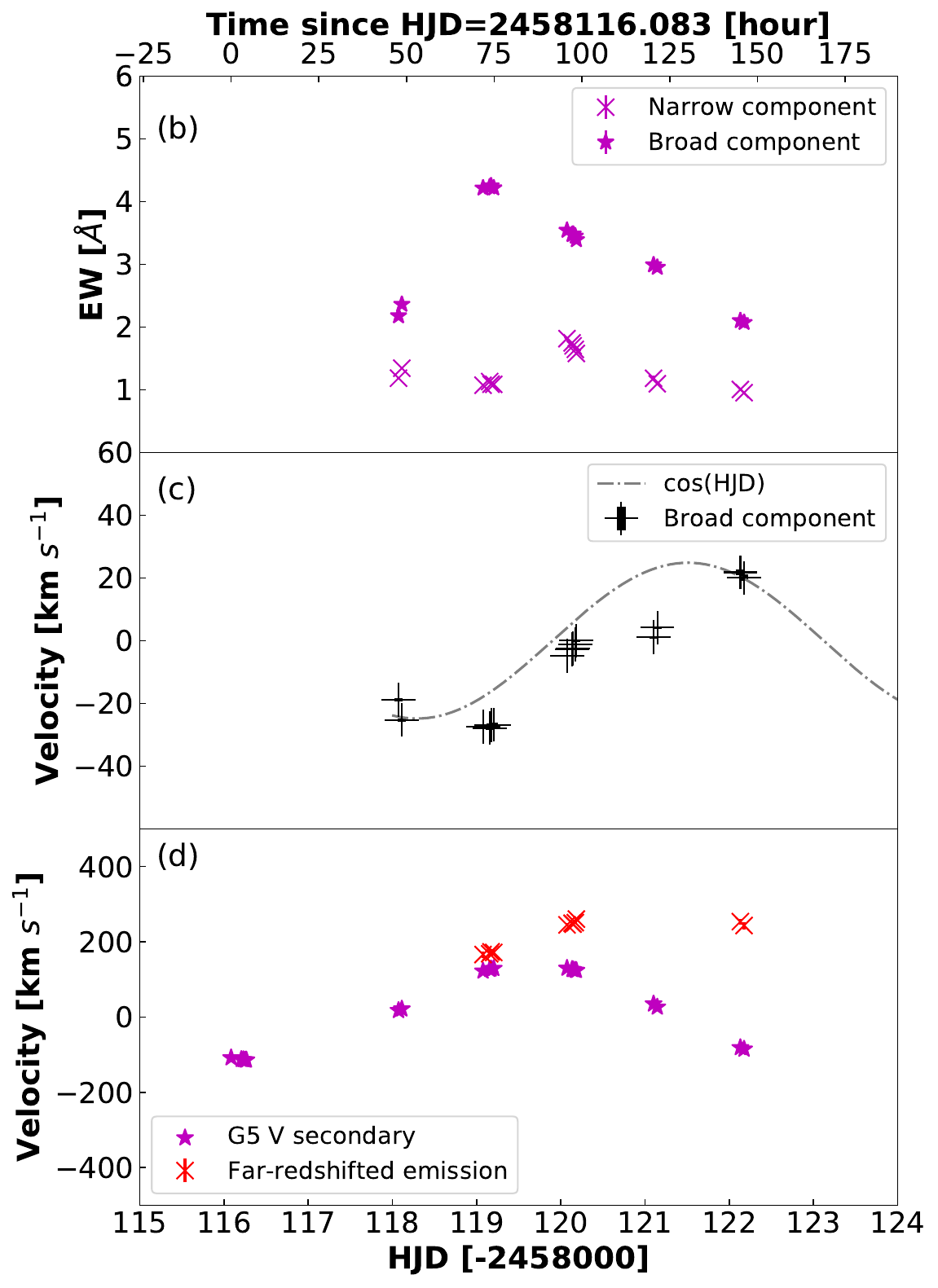}
\caption{(a) Examples of Gaussian fitting for the H$_{\alpha}$ subtractions. Left part: Observed (black solid lines) and synthesized (magenta dashed lines) H$_{\alpha}$ spectra. Observing dates and phases are also marked. Right part: Gaussian fitting (color dashed lines) for the H$_{\alpha}$ subtractions (black solid lines). All subtracted spectra are corrected to the rest velocity frame of the K0~IV primary star of UX~Ari. The insets show zooms into the spectral ranges with the far-redshifted emission components. (b) Temporal variations of the EWs of the narrow and broad emission components. (c) Temporal variations of the velocities of the broad emission component and the fitting line. (d) Temporal variations of the velocities of the far-redshifted emission component and the G5~V secondary component relative to the K0~IV primary component.}
\label{Fige1}
\end{figure*}
%-------------------------------------------------------------------------------------------------------------------------%
\begin{figure*}
\centering
\includegraphics[width=19.0cm,height=9.5cm]{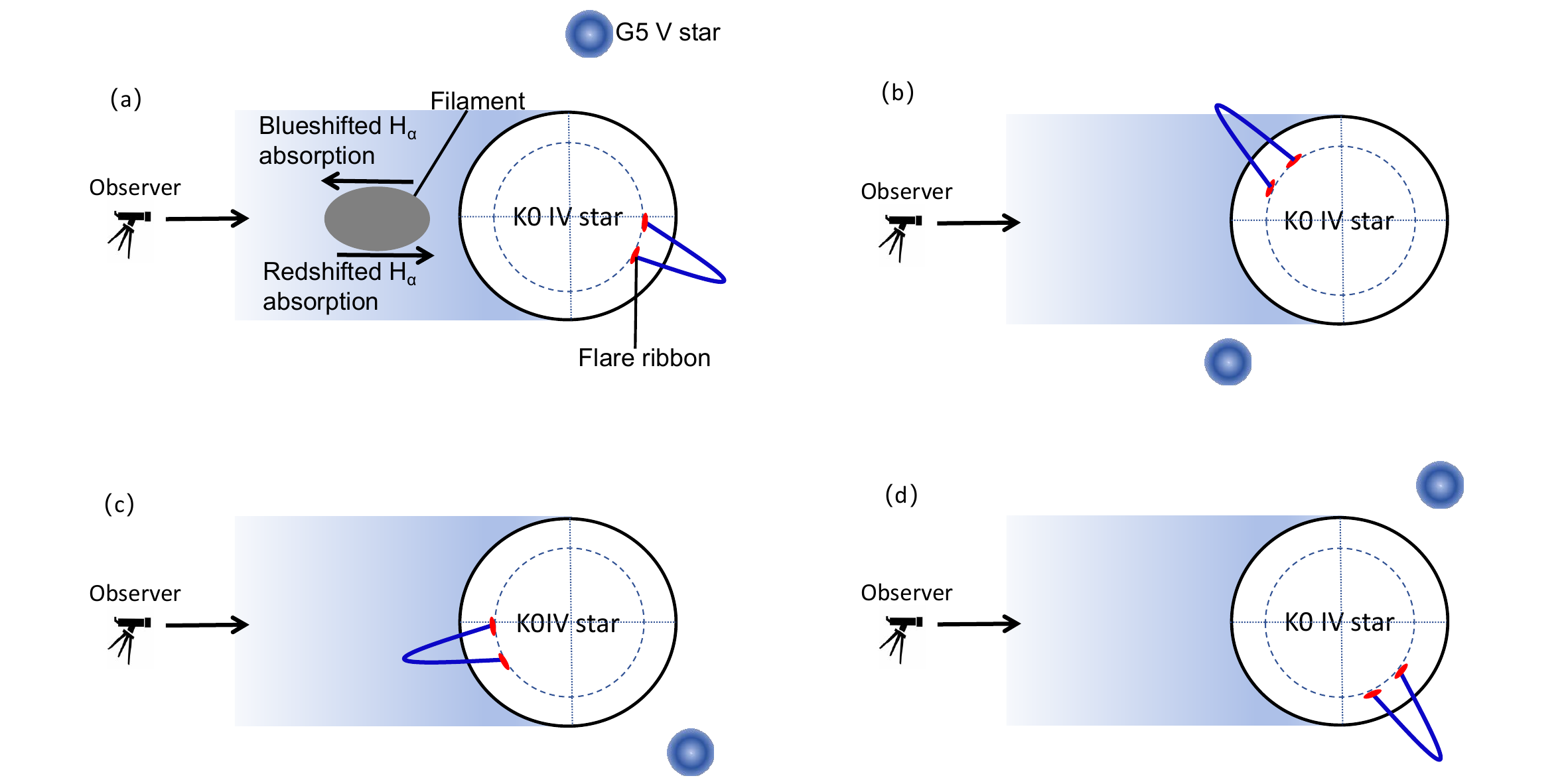}
\caption{Schematic diagram that expresses the interpretation of the Doppler-shifted excess absorptions and the possible flaring location. Panel~(a) represents the occurrence of the Doppler-shifted absorption features observed on 2017 December~28 and the flaring location at that time. Panels (b), (c) and (d) depict the possible flaring location on 2017 December~31, 2018 January~1, and 2018 January~3, respectively. The dashed circles represent the latitude of 50$^{\circ}$. The small telescope illustration is obtained from Figure~7 of \citet{Inoue2023}.}
\label{Fige2}
\end{figure*}
%-------------------------------------------------------------------------------------------------------------------------%
For the other flaring spectra observed from 2017~December~30 to 2018~January~3, the subtracted H$_{\alpha}$ profiles are also modelled using the Gaussian fitting method. The H$_{\alpha}$ profiles usually can be well fitted by a narrow emission component and a broad emission one. As displayed in Fig.~\ref{Fige1}~(a), one spectrum from each night's observation is presented as an example. In Fig.~\ref{Fige1}~(b) and (c), the EWs of the narrow and broad emission components and  the velocities of the broad emission component are respectively plotted against HJD and flare time. It can be seen that the broad emission component is dominant and Doppler-shifted. Additionally, for the observations on 2017 December~31, 2018 January~1 and 3, there are far-redshifted emission components in the fitted profiles. These components exhibit bulk velocities of about 170~km~s$^{-1}$, 250~km~s$^{-1}$, and 250~km~s$^{-1}$, respectively. The velocities of the far-redshifted emission components are plotted against HJD and flare time in Fig.~\ref{Fige1}~(d), and the velocities of the G5~V secondary component relative to the K0~IV primary component are also plotted. It can be found that these far-redshifted emission components are not associated with the G5~V secondary component (see Fig.~\ref{Fige1}~(d)). Furthermore, the flaring spectra exhibited broader profiles compared to the subtracted pre-flare reference spectra (see Fig.~\ref{Figb1}). Consequently, these far-redshifted emissions can be attributed to plasma motions associated with the flare event.

As discussed in Section 4.2, the flare region was located at the stellar limb that is moving towards us at phase~0.045. Taking further account of the orbital motion of UX~Ari, the flaring location at a different orbital phase can be inferred and a schematic diagram illustrating the potential flaring location is presented in Fig.~\ref{Fige2}. For the observation on 2017 December~28, we can infer that the flare was located at the hemisphere nearly opposite to that of the filament eruption (Fig.~\ref{Fige2} (a)). By 2017 December 31, the star had rotated approximately 44$^{\circ}$ from phase~0.045. Consequently, the flare region shifted to a location similar to that shown in Fig.~\ref{Fige2} (b), and this suggests that the far-redshifted emission component indicates the occurrence of the downward-moving plasmas toward the star. Subsequently, by 2018 January~1, another rotation of about 56$^{\circ}$ had taken place and correspondingly led to movement of the flare ribbon (see Fig.~\ref{Fige2} (c)). The far-redshifted emission component still represents the occurrence of the downward-moving plasmas. The downward-moving plasmas can be attributed to chromospheric condensation, flare-driven coronal rain, or material falling in a prominence eruption based on current knowledge of solar flares and observing facts on stellar flares. Finally, by 2018 January~3, an additional rotation of about 110$^{\circ}$ had occurred, resulting in a corresponding shift in the position of the flare ribbon as illustrated in Fig.~\ref{Fige2}~(d). At this point, the far-redshifted emission component implies the presence of the upward-moving plasmas from the stellar surface, which might be in the form of chromospheric evaporation or prominence eruptions. A number of studies had reported and provided detailed analyses of comparable phenomena in a variety of active stars. \citep[e.g.][]{Fuhrmeister2018, Vida2019, Wu2022, Cao2024}.

Based on the method initially introduced by \citet{Houdebine1990} and recently used by several authors \citep[e.g.][]{Koller2021, Wang2022, Wu2022, Cao2024}, the minimum masses of the moving plasmas can be estimated from the Doppler-shifted emission components. The calculation strategy can be found in \citet{Cao2024}, and therefore the minimum masses are derived to be $2.3 \times 10^{20}$~g for the observation on 2017 December~31, $8.5 \times 10^{19}$~g on 2018 January~1, and $2.3 \times 10^{20}$~g on 2018 January~3, respectively. The estimations are also consistent with the values of the possible erupted prominences or moving plasmas on other stars.

Moreover, for the observations obtained on 2017 December~30 and 2018 January~2, there are no similar Doppler-shifted emission features presented in the H$_{\alpha}$ spectra. In addition to real absence of such features, one plausible explanation is that the flare ribbons are likely located at positions where their radial orientations are nearly perpendicular to line of sight during these observations. Consequently, even if the moving plasmas were occurring during these observations, they could not be detected.
%
%%%%%%%%%%%%%%%%%%%%%%%%%%%%%%%%%%%%%%%%%%%%%%%%%%%%%%%%
\section{Estimation of the filament mass and length scale}\label{secf}
The mass of the filament is simply calculated by using the Becker cloud model \citep{Mein1988}, which had been used by \citet{Namekata2022} in their analysis of EK~Dra. The same parameter assumptions as \citet{Namekata2022} are used, including optical depth at the line centre of the ejected plasma $\tau_{0}$ of 5, local plasma dispersion velocity $W$ of 20~km~s$^{-1}$, and source function of the ejected plasma $S$ of 0.1. These parameter assumptions were derived from solar observations and more details can be found in \citet{Namekata2022}. The surface area of the filament $A_{F}$ is obtained as $1.5~\times~10^{23}$~cm$^{2}$ ($\sim$~\%32 of the stellar disk). Following \citet{Namekata2022} and \citet{Namekata2024}, the length scale of the filament $L_{F}$ can be obtained by assuming that the filament has a ``cubic''-like structure ($L_{F}$~=~($A_{F}$)$^{0.5}$) for the lower limit and a ``cylinder''-like structure ($L_{F}$~=~(10$A_{F}$)$^{0.5}$) for the upper limit. Therefore, the length scale of the filament  $L_{F}$ is estimated to be $3.9-12.3~\times~10^{11}$~cm. Next, based on the plasma model, the hydrogen column density is derived as $3.4-5.9~\times~10^{21}$~cm$^{-2}$ from the assumed optical depth and the calculated length scale of the filament. Therefore, the filament mass is derived as $8.5-15.0\times~10^{20}$~g by multiplying the hydrogen column density by the filament area. 
%
%-------------------------------------------------------------------------------------------------------------------------%
\end{appendix}
\end{document}